\pdfoutput=1

\documentclass[sigplan,nonacm]{acmart}

\makeatletter
\renewcommand{\@ACM@checkaffil}{%
  \if@ACM@instpresent\else
    \ClassWarningNoLine{\@classname}{No institution present for an affiliation}%
  \fi}
\makeatother

\usepackage{array}
\usepackage{pifont}
\usepackage{xurl}
\usepackage{xspace}

\graphicspath{{figures/}}

\definecolor{markyes}{HTML}{2E7D32}
\definecolor{markno}{HTML}{C0392B}
\newcommand{\yes}{\textcolor{markyes}{\ding{51}}}
\newcommand{\no}{\textcolor{markno}{\ding{55}}}
\newcommand{\hd}[2]{\shortstack[c]{#1\\#2}}

\newcommand{\sysname}{Grouper}
\newcommand{\sys}{\sysname\xspace}

\newcommand{\us}{\ensuremath{\mu}s\xspace}
\newcommand{\uT}{\ensuremath{\mu}Tune\xspace}

\newcommand{\cbrk}{\discretionary{-}{}{}}
\newcommand{\hotelres}{\texttt{ho\cbrk tel\cbrk Res\cbrk er\cbrk va\cbrk tion}\xspace}
\newcommand{\browseprod}{\texttt{browse\cbrk Prod\cbrk uct}\xspace}
\newcommand{\socialnet}{\texttt{so\cbrk cial\cbrk Net\cbrk work}\xspace}

\begin{document}

\title[\sysname]{\sysname: Scheduling Groups for Multi-Tenant
  Microsecond-Scale Microservices}

\author{Koosha Kazemi}
\affiliation{%
  \institution{Sharif University of Technology}
}
\email{koosha.kazemi24@sharif.edu}

\author{Mohammad Siavashi}
\authornote{Work done while at Iran University of Science and Technology.}
\affiliation{%
  \institution{KTH Royal Institute of Technology}
}
\email{siavashi@kth.se}

\author{Ahmad Siavashi}
\authornote{Work done while at Amirkabir University of Technology.}
\affiliation{%
  \institution{Independent Researcher}
}
\email{a.siavashi@outlook.com}

\author{Mohammad Izadi}
\affiliation{%
  \institution{Sharif University of Technology}
}
\email{izadi@sharif.edu}

\renewcommand{\shortauthors}{Kazemi et al.}

\begin{abstract}
Microsecond-scale core allocation makes colocating latency-critical services
with batch work worthwhile. A thread that finds no work parks within
microseconds and its core goes to a batch task. Putting one back costs
${\sim}$18\,\us, as the allocator must discover that a core is wanted and
then take it from the batch task holding it. A monolith pays that tax once per
request, a microservice chain pays it at every hop in both directions, and a
multi-tenant host multiplies it again, because every tenant's hops queue at the
same allocator. On our port of DeathStarBench's \hotelres, going from two
tenants to ten takes a hop from 39 to 222\,\us and a 10-RPC path's median from
456 to 2,445\,\us, a fivefold degradation even though no tenant's own load
changed.

We introduce \sys and the \emph{scheduling group}, a set of isolated runtimes
that the allocator treats as one allocation and accounting unit, whose members
may hand cores directly to one another. A service sending an RPC donates its core
to the peer through
an unprivileged kernel fast path, so the core follows the request through the
call graph. The allocator retains control through reconciliation, core-addressed
revocation and a pooled budget but leaves the critical path; its load falls
from $\Theta(R{\cdot}H)$ to $\Theta(R)$ in request rate $R$ and hop count $H$.
Over a grid of two to ten tenants at 1,000--30,000 requests per second each,
\sys outperforms Caladan (the allocator Junction also builds on) and Linux by up
to 7.9$\times$ and 3.4$\times$ at the median and 4.1$\times$ and 14.2$\times$
at the tail, and leaves batch work more throughput than Caladan at over 70\% of
load points.
\end{abstract}




\maketitle

\section{Introduction}
\label{sec:intro}

\begin{figure}[t]
  \centering
  \includegraphics[width=\columnwidth]{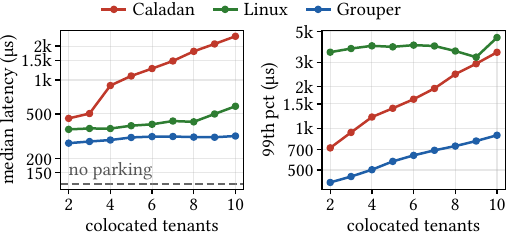}
  \caption{\looseness=-1\textbf{Adding neighbours, not load.} Nine-service \hotelres
    deployments share one 72-core machine with a batch task, each offered a
    fixed 20,000 requests per second. From left to right, no tenant's call
    graph, hop count, per-hop compute or request rate changes, only how many
    other tenants share the machine. Median (left) and 99th percentile
    (right) of the 10-RPC \texttt{search} path. The blue curve in both panels
    is \sys. Note the log axes.}
  \label{fig:latency}
\end{figure}

Datacenter operators want latency-critical (LC) services that respond in
microseconds~\cite{barroso2017killer}, and enough batch work colocated with them
that the machines are not mostly
idle~\cite{lo2015heracles,chen2019parties,iorgulescu2018perfiso}. Those goals
pull against each other. The
line of work that made this practical (IX~\cite{belay2014ix},
ZygOS~\cite{prekas2017zygos}, Shinjuku~\cite{kaffes2019shinjuku},
Shenango~\cite{ousterhout2019shenango}, Caladan~\cite{fried2020caladan},
Junction~\cite{fried2024junction}) moves
core allocation out of the kernel scheduler into a dedicated userspace
allocator that reallocates far faster than Linux can, harvesting a core as soon
as its owner runs out of work and handing it to a best-effort (BE) task. Every
one of them places a core \textbf{by observation}. The allocator polls,
discovers that some application is congested, and moves a core to it, which is
the only option available when the applications are unrelated.

Harvested cores pay a ``parking tax''. Waking a runtime whose threads are all
parked is a distributed operation costing ${\sim}$18\,\us when a BE task
holds the core and must be made to cede.

A monolithic server pays it once per request. A core is granted on arrival and
the request keeps it for its whole lifetime, calling functions sequentially
inside one address space. A microservice deployment pays it at each \emph{hop}, and
on both legs of each hop. By the time a reply comes back the caller has exhausted
its spin window and parked, so the reply must wake it again, while the compute
at each stop is often smaller than the cost of arriving there, a sub-microsecond
hash lookup or a field rewrite. The regime in which this hurts is the one
production runs in. Microservices are provisioned with headroom and run at low
to moderate utilization, where inter-arrival spacing exceeds the spin
window.

The penalty extends to processor cache locality. Allocator-mediated schedulers
blindly harvest cores mid-request when a thread yields: an intervening batch task
evicts cache warmth, while later wakes frequently land on another core,
incurring cold misses and cross-core coherence traffic on the request's
critical path.

\textbf{Multi-tenancy is what turns the tax into a collapse.}
Figure~\ref{fig:latency} shows it, with the same per-tenant load throughout and
only the number of deployments sharing the machine increasing. Over that sweep the
median of a 10-RPC path grows 5.4$\times$ and its 99th percentile 4.9$\times$.
The dashed line is what that same path costs when kthreads never park; the
difference is the machine failing to put a core where the request already is.
Observation is centralized. One allocator must rank every
runtime to decide which deserves a core, so its reaction time is set by how many
runtimes there are, not by how much work they have. As services multiply, the
scan that produces the ranking stops fitting in its own period and runs on every
iteration, stretching the loop every grant waits on.

\textbf{Most of that discovery is unnecessary.} The information the allocator
polls for is already in the application's dataflow, and only the application has
it. A service that has finished its part of a request and is sending an RPC
knows which peer needs a core next, and that it is itself about to stop needing
the one it is standing on. If the two processes are mutually trusted, the sender
can hand its core over, and there is nothing left to discover.

\sys builds on that. A \textbf{scheduling group} is a set of isolated
runtimes, with separate address spaces, failure domains and binaries, that the
allocator treats as one allocation and accounting unit, and whose members
are mutually authorized to hand each other cores through an unprivileged kernel
fast path in about 4\,\us, with no allocator round trip. The allocator is
involved once per request, at the edge where the request enters the group; every
interior hop is a transfer it never sees.

Making that safe is most of the work. Two things break as cores move without
the allocator's knowledge: accounting drifts, and thread-targeted revocation
preempts the wrong occupant. \sys responds with per-iteration reconciliation
and core-addressed revocation. Direct handoff also avoids NIC round trips via a
shared-memory datapath delivered synchronously with the donation, leaving an
interior hop free of receive queues, hashes or steering decisions.
Because execution stays on the same physical core, the handoff acts as a
cache-preserving execution migration, retaining processor cache and translation
locality across process boundaries.

A group that passes one core along a call chain makes the
\emph{request}, not the task, the unit of host scheduling, and we state that as
an invariant. \textbf{At every instant, a request in flight inside a scheduling
group occupies exactly one core, and no other request occupies it.} Under it
kthreads stop being interchangeable workers that steal from one another's
runqueues and become deterministic landing \emph{pads} for specific request
\emph{lanes}, which gives the chain the locality of a single-threaded monolith,
gives the group an explicit concurrency bound, and yields edge admission
control.

We contribute the scheduling group abstraction and its trust model
(\S\ref{sec:design-group}); the one-core-per-request invariant and the
group-wide lane naming that makes it enforceable across address spaces and
yields edge admission control (\S\ref{sec:design-invariant});
unprivileged directed handoff together with the reconciliation that keeps a
central allocator's accounting exact
(\S\ref{sec:design-handoff}--\S\ref{sec:design-reconcile}); and an evaluation
of \sys versus Caladan and Linux.

\section{Background and Motivation}
\label{sec:background}

\subsection{Core harvesting and the parking tax}
\label{sec:bg-harvest}

A core-allocating host scheduler divides a machine between a privileged control
plane and unprivileged runtimes. The \textbf{allocator} owns the cores. It
interleaves a sub-microsecond fast pass with a periodic \textbf{slow pass} that
scans every runtime's queues to estimate congestion and decide allocations, and
grants or revokes a core through a kernel module that wakes a specific thread
on a specific core. Each runtime is a userspace scheduler running user-level
threads (\emph{uthreads}) on kernel threads (\emph{kthreads}), one per granted
core~\cite{anderson1991activations,qin2018arachne}. The structure is common to
this class; we measure
Caladan~\cite{fried2020caladan}, whose slow pass is scheduled every 10\,\us.

The harvesting rule is aggressive. A kthread that finds its runqueue, ingress
queue and timer wheel empty spins for 2\,\us and then parks, which
returns a core to batch work almost as soon as it is genuinely idle, and means
a service idle for even a few tens of microseconds between requests is reliably
found with all of its kthreads asleep. Waking one back up is a distributed
operation. The allocator must \emph{notice} the work; rank congested runtimes
and choose a core; if a batch task occupies that core, interrupt it and wait for
it to cede; and only then wake the target kthread there.
Table~\ref{tab:wake} breaks the parking tax down.

\begin{table}[t]
  \caption{Where a wake goes (\us, p50). One hop, 70 schedulable cores, vfio
    directpath, x264 on every core the service does not hold.}
  \label{tab:wake}
  \begin{tabular*}{\columnwidth}{@{}@{\extracolsep{\fill}}lcc@{}}
    \toprule
    stage & idle core & BE on the core \\
    \midrule
    detect arrival             & 0.1 & 0.2 \\
    choose core                & 0.4 & 0.5 \\
    IPI + signal               & 0.0 & 5.7 \\
    cede                       & 0.0 & 1.3 \\
    context switch             & 5.6 & 9.9 \\
    \midrule
    \textbf{total}             & \textbf{6.1} & \textbf{17.6} \\
    \bottomrule
  \end{tabular*}
\end{table}

\subsection{The tax is per hop, and its price is set by the neighbours}
\label{sec:bg-perhop}

A monolithic server amortizes one wake over an entire request. A microservice
chain does not. Each RPC crosses a process boundary, and each crossing finds a
service whose kthreads have parked, on the way down because the callee has
been idle since its last request, and on the way back because the caller parked
while blocked on the reply. A chain that makes $H$ RPCs therefore incurs roughly
$2H$ wakes.

We ported DeathStarBench's \hotelres~\cite{gan2019dsb} onto
such a runtime to measure this on an actual call graph. Its \texttt{search} path is
four levels deep (client $\rightarrow$ frontend $\rightarrow$ search
$\rightarrow$ rate $\rightarrow$ kv) for a total of
\textbf{10 RPCs}, and non-empty searches are ${\sim}$31\% of the standard mix.
Weighting the whole mix gives \textbf{$\approx$5.6 RPCs per request}, and the
compute at each
stop is often a single hash lookup and a field copy, well under a microsecond.
This is a workload where the scheduler, not the application, sets the
latency~\cite{sriraman2018utune}.

That shape is representative. Production traces from Alibaba~\cite{luo2021alibaba}
find an average call-graph depth of 4.27, with many deep graphs reducing to a
single long chain; Google's Online Boutique, which \S\ref{sec:eval-depth}
measures, is a shipped instance of that tail, 11.8 RPCs from a graph three
levels deep. Every level is a boundary crossing that finds a parked service,
which is why even modest depth pays a large
tax~\cite{dean2013tail,zhu2023meshinsight}.

\textbf{Scheduling is most of this latency, and
what raises it is the neighbours rather than the load}
(Figure~\ref{fig:latency}). Six times the offered load per tenant costs the
allocator 1.7$\times$; holding each tenant's rate fixed and adding five times
as many tenants costs it 5.4$\times$. Nothing about any individual request has
changed. What changes is the allocator's reaction time. With dozens of colocated
services, the allocator's slow pass must inspect hundreds of kthread queues;
when this scan exceeds 10\,\us, it stops being amortized and
runs on \emph{every} loop iteration. The dataplane loop period blows up from
sub-microsecond times to tens of microseconds, delaying every core grant and
making per-hop delivery surge from 38.6\,\us at two tenants to 221.6\,\us at
ten (\S\ref{sec:eval-hop}). Making the polling cheaper does not fix this
(\S\ref{sec:bg-remedies}). \textbf{The bottleneck is the per-hop acquisition
itself}, whose decision loop stretches with every tenant on the machine.

\subsection{Why the obvious remedies fail}
\label{sec:bg-remedies}

\emph{Spin longer.} A runtime can spin longer before
parking~\cite{karsten2020threading}, or pin a kthread that never parks. Both
shrink that tax, and pinning removes it outright, but both charge it to the same
account. A spinning core is held whether or not a request arrives, and every one
of them is a core the batch task cannot have. The reservation is per service, so
even a small microservice deployment fits few tenants on a machine and leaves
the batch task very little. \S\ref{sec:eval-price} prices the whole sweep.

\emph{Make polling cheaper.} Junction~\cite{fried2024junction} scales the
allocator to thousands of instances with a NIC event queue that arms idle
receive queues instead of polling them and a 16\,\us hierarchical timer wheel
that skips idle runtimes. This helps at lower load but converges on the default
as the machine fills. Scanning less often would help too, but the slow
pass supplies the signal that ranks runtimes for a core, the estimates that
decide who may share one and the queueing delays the policy compares, so
widening its interval coarsens all of them at once, precisely when tenants are
most numerous and interference mitigation matters most. The hops have to stop arriving at that scan.

\begin{table*}[t]
  \caption{\textbf{What each system offers a microservice chain.}}
  \label{tab:systems}
  \small
  \setlength{\tabcolsep}{4pt}
  \begin{tabular*}{\textwidth}{@{}@{\extracolsep{\fill}}lccccccrr@{}}
    \toprule
    System
      & \hd{Reclaims a core}{from batch in \us}
      & \hd{Interference}{control}
      & \hd{Interior hop}{off the NIC}
      & \hd{Hop cost flat}{in tenants}
      & \hd{Core follows}{the request}
      & \hd{Admission}{control}
      & \hd{search p50}{(\us)}
      & \hd{search p99}{(ms)} \\
    \midrule
    Linux              & \no  & \no  & \yes & \yes & \no  & \no  & 363--1,057 & 3.3--11.0 \\
    Caladan / Junction & \yes & \yes & \no  & \no  & \no  & \no  & 447--2,565 & 0.58--3.86 \\
    \sys             & \yes & \yes & \yes & \yes & \yes & \yes & 271--371   & 0.37--1.02 \\
    \bottomrule
  \end{tabular*}
\end{table*}

Table~\ref{tab:systems} is where this leaves the three systems we measure, and
the two baselines fail in opposite ways. Linux has no central allocator,
so an interior hop costs a tenant the same whatever its neighbours do, but it
cannot take a core back from a batch task in microseconds and pays for that in
the tail. Caladan (and Junction, which vendors the same allocator) can, but
every hop is then a request to one machine-wide decision maker whose reaction
time is set by how many runtimes are asking. Only \sys provides both.

\section{Design}
\label{sec:design}

\begin{figure}[t]
  \centering
  \includegraphics[width=\columnwidth]{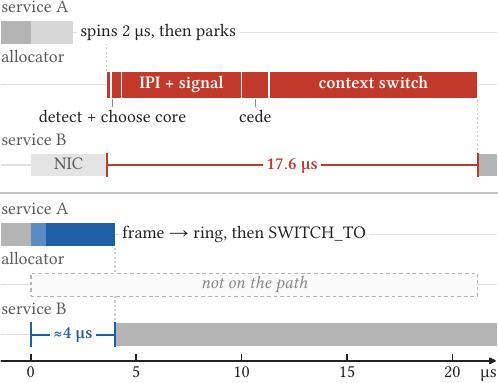}
  \caption{\textbf{The cost of one hop, with and without the allocator.} Park-and-wake
    on top, with Table~\ref{tab:wake}'s stages in proportion. Directed handoff
    into the receiver's ring on the bottom.}
  \label{fig:hop}
\end{figure}

\begin{figure}[t]
  \centering
  \includegraphics[width=\columnwidth]{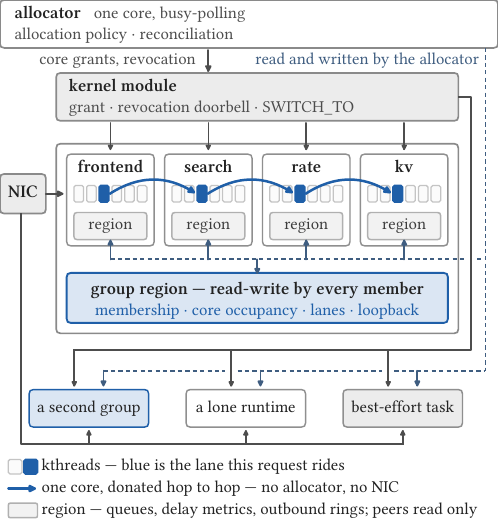}
  \caption{\textbf{\sys's system architecture.} Isolated members share a
    pooled core budget and one group region. An interior hop leaves the frame
    in the ring its peer drains and donates the core, so neither the allocator
    nor the NIC is on its path. The allocator is the fallback for any hop that
    is not donated. The same allocator and the same NIC serve the rest of the
    machine (another group, an ungrouped runtime, a best-effort task).}
  \label{fig:arch}
\end{figure}

A scheduling group is a set of $N$ runtimes that declare, in their
configuration files, that they belong to the same named group. The allocator
treats the group as one allocation and accounting unit, with one pooled core
budget and one shared region of state, and authorizes every member's kernel
threads to hand a core directly to any other member's. Members remain separate
processes with separate address spaces, failure domains and binaries, and
nothing about the application programming model changes. The design must do
four things at once. It must move a core between address spaces in microseconds
with no allocator round trip (\S\ref{sec:design-handoff}); keep the allocator's
accounting exact and its preemption working when it no longer knows who
occupies a core (\S\ref{sec:design-reconcile}, \S\ref{sec:design-revoke}); keep
process isolation, so that no member holds a writable mapping of another's
memory; and fall back to park-and-wake whenever group state is absent or
inconsistent (\S\ref{sec:design-group}). Figure~\ref{fig:hop} contrasts a hop
with and without the allocator on its path; Figure~\ref{fig:arch} shows the
resulting architecture.

\subsection{Trust, lifecycle and the fallback rule}
\label{sec:design-group}

\paragraph{Trust model} A group encapsulates \emph{one tenant's application},
the same team's services, deployed together, already able to invoke each other.
It never spans tenants and is not a security boundary between mutually
distrusting parties, which is what makes mutual donation sound. Isolation is
preserved across process boundaries. Loopback rings are mapped asymmetrically
and read-only, and handoff authority is granted by the privileged allocator
rather than claimed by members. A member's authority also stops at the group. A
donation moves a core the group already holds, the pooled budget is enforced by
the allocator when it grants (\S\ref{sec:design-policy}), and a member that
will not yield is preempted by the same core-addressed revocation as any other
occupant (\S\ref{sec:design-revoke}), so nothing a member does reaches another
tenant. A group is host-local as well. A deployment spanning machines runs one
group per host, with ordinary networking between them.

\paragraph{Lifecycle} A group is formed, made ready, and dissolved by the
allocator, and outside that window its members are ordinary runtimes. Two
configuration lines make a runtime a member, a group name and an expected member
count $N$. The allocator allocates a shared region on first registration and
passes it to each member as it attaches; until all $N$ have arrived the group is
not \texttt{ready}. When the last registers, the allocator authorizes every
member thread in the kernel module, publishes the lane count
(\S\ref{sec:design-invariant}) and sets \texttt{ready}. A member that exits or
crashes dissolves the group rather than repairing it, and the survivors fall
back to park-and-wake, with no error path.

\paragraph{The fallback rule} Nothing in the shared region is needed for
correctness. Every path that consults it (donation, lane acquisition,
loopback transmission, reconciliation) checks readiness first and takes the
park-and-wake path if anything is missing or inconsistent. That is what makes
group state a \emph{hint layer} over an unmodified allocator, and why a
misbehaving or crashed member costs performance rather than correctness.

\subsection{Directed handoff}
\label{sec:design-handoff}

The fast path is the unprivileged \texttt{SWITCH\_TO} ioctl. A
kthread owning a core calls it with a target thread ID, and the kernel module
hands the core over. After confirming caller occupancy and that no allocator
revocation is pending, it wakes the target, transfers the occupancy record, and
yields. Linux then switches directly to the successor on the same physical core
(about 4\,\us), keeping the core's busy flag set throughout so it never appears
idle. Because execution stays on the same core, this migration preserves warm
L1/L2 caches, while Linux's PCID support retains TLB translations across the
address-space switch without a full flush. A refused donation is not an error;
the caller falls back to ordinary park-and-wake.

\subsection{Reconciliation}
\label{sec:design-reconcile}

Handoffs happen inside groups and the allocator never sees them, so its record
of who sits on which core goes stale in both directions at once. It still counts
the donor as holding the core it gave away, and does not count the recipient as
holding anything. Every allocation decision is computed from those counts, so
they have to be accurate. Each core's record in the shared region carries a handoff
sequence number and the allocator caches a pointer to it, so discovery costs one
load per core and is almost always an equality; when it differs, the allocator
moves the active-thread counts across. It is discovery and not enforcement.
What it cannot resolve, a thread already active elsewhere, because handoffs chain,
is left a pass, and a periodic repair concedes any lasting disagreement back
to the runtime, which is the authority on whether its own thread is running.

\subsection{One core per request}
\label{sec:design-invariant}

\begin{quote}
\bfseries At every instant, a request in flight inside a scheduling group
occupies exactly one core, and no other request occupies it.
\end{quote}

The invariant restores, to a distributed call chain, the property a monolith gets
for free. A request arrives on a core and keeps it until it leaves, because the
calls it makes are function calls. \sys makes the same true across process
boundaries, so kthreads cease being interchangeable workers and become
deterministic landing pads for particular requests.

\begin{figure}[t]
  \centering
  \includegraphics[width=\columnwidth]{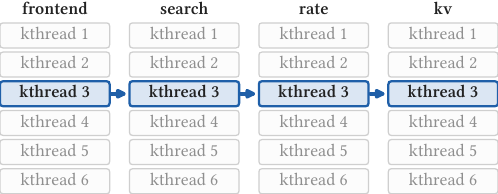}
  \caption{\textbf{A request's lane along the call graph.} Lane $l$ is kthread
    $l$ at every member; the core travels with the request. The request is shown
    going through lane 3; a mis-steered packet claims a free lane at the first
    hop.}
  \label{fig:lane}
\end{figure}

\paragraph{Lanes} A \emph{lane} is the group-wide name for the core a request is
riding. Lane $l$ means kthread $l$ at every member. A request holding it runs on
kthread $l$ wherever it is, sends into ring $l$ of whatever peer it calls, and
donates to kthread $l$ (Figure~\ref{fig:lane}). Because the name must mean the
same thing everywhere, every member's kthread count is the same.

Changing lanes is not allowed, so \textbf{lanes cannot be shared}. Two requests
riding lane $l$ would be on kthread $l$ of every service they both reached,
which is the double occupancy the invariant forbids. A lane is therefore
\emph{booked} on entry and cleared on exit. Two consequences carry the rest of
the design. The lane count is a hard concurrency bound, which
\S\ref{sec:design-admission} turns into admission control; and because only the
holder of lane $l$ can send along it, a kthread's supplier is unambiguous with
no arbitration. Booking occurs only at the edge. Inside the group a frame
arrives through the ring that already names its lane.

\paragraph{Fixed receive queues} A general dataplane must re-steer traffic
whenever kthreads park or wake, forcing awake threads to adopt parked
queues
and churning flow assignments or NIC RSS
tables~\cite{barbette2019rsspp,pesterev2012affinity,katsikas2018metron}. This
adds synchronization
overhead on core transitions and breaks request locality.
\sys eliminates steering altogether by permanently binding receive queue $i$ to
kthread $i$. Because the core travels along the request's lane, the core moves
to the work rather than steering work onto an awake core
(\S\ref{sec:design-loopback}).

\paragraph{First-hop placement} Inbound network packets are steered by NIC RSS,
which is unaware of the group's lane bookings. If a packet lands on a kthread whose
lane is already booked, the runtime claims an available parked lane, places the
request onto that lane's kthread, and donates the arrival core directly to it.

\paragraph{Transparent integration} Applications already choose per-core
resources (connection pools, client handles, sharded buffers) through an
affinity call returning the current kthread index. Inside a group that call
returns the current \emph{lane} instead, so connection $l$ is chosen, its frames
go to ring $l$ of the peer, and that ring is drained by the peer's kthread $l$,
the kthread the core was donated to. Unmodified applications keep flow
affinity without a line of change.

\subsubsection{Admission control at the edge}
\label{sec:design-admission}

When every lane is held, booking fails, and the runtime raises a flag that the
application reads once and turns into a protocol-level refusal (an HTTP 503,
say). One hook in the edge gateway is the whole application-side cost; the
runtime cannot render the refusal itself, because its only means of refusing is
to abort the connection, which on a pooled connection destroys every other
request sharing it just to turn away one.

A failed booking means the group already has as many requests in flight as it
has lanes, and the honest answer is to refuse that one rather than run it on a
core another request is riding. A request admitted into an oversubscribed group
accumulates latency at every hop and very likely misses its tail SLO
\emph{after} burning several hops of
work~\cite{cho2020breakwater,welsh2001seda}, while the cores it would have
consumed keep running batch work if it is turned away. It can also make a
laneless request \emph{wait} instead; this is off by default.

\subsection{Group loopback}
\label{sec:design-loopback}

Microservices that talk over the NIC pay double PCIe traversals, DMA and
packetization~\cite{kalia2019erpc,li2019socksdirect}, and compete with external
traffic for NIC resources. \sys replaces intra-group networking with a
shared-memory loopback transport that delivers messages through cache and
memory synchronously with the core handoff.

Each ordered pair of members gets one lockless channel per receiving kthread.
The descriptor ring lives in the \emph{sender's} region and the consumer
writeback in the \emph{receiver's}, so each side writes only its own memory and
reads the other's read-only. No member holds a writable mapping of another's
address space, as a property of the layout rather than an enforced rule.
Because a ring names a (member, kthread) pair, the sender chooses which of the
receiver's kthreads picks up the frame with no RSS hash or flow table.
That is sound only because the sender then donates its core to that kthread.
The send writes the descriptor and pending hint \emph{before} donating, so the
frame is in the ring when the receiver resumes. Because handoff runs the
receiver on the same physical core, this ``hot handoff'' serves descriptors and
payload directly from local L1/L2 cache. By contrast, allocator-mediated wakes
often schedule the receiver on another core or after an intervening batch task,
incurring cross-core coherence traffic and cold cache misses. The receiver copies
the frame into its address space; the layout also admits zero-copy for larger
transfers.

That hint is the recovery path as well. Loopback bypasses hardware NIC queues, so a
frame could sit unnoticed if a donation is refused or preempted. The allocator
sweeps the hints on its slow pass (\S\ref{sec:design-cost}) and wakes any
recipient still parked that has frames pending.

\subsection{Core-addressed revocation}
\label{sec:design-revoke}

An allocator revokes a core by writing a sequence number into the queue pointers
of the kthread it believes is there, which it can do, because it put them
there. With groups, it cannot. Cores change hands without its involvement, so a
revocation addressed to a thread reaches the wrong one. \sys addresses
revocation to the \textbf{physical core} instead. The allocator increments a
cede sequence in the core's record; whoever is executing there notices the
mismatch at its next preemption check, yields, and acknowledges by matching the
sequence. An arriving kthread also retires any outstanding cede on the core it
\emph{lands} on, so a stale revocation cannot immediately re-park whoever just
arrived. Because the revocation is durable state rather than a message, the
interrupt that makes a compute-bound occupant look at it is only a doorbell.
A lost doorbell costs latency, never correctness, and can be rung again without
rewriting anything that names a kthread.

\subsection{Allocation policy}
\label{sec:design-policy}

\sys's allocation policy ranks runtimes by how badly each needs a core and
enforces a configured core limit and guarantee for each.

\paragraph{Pooled budget} A group is one entity for provisioning with its own
configured limits and guarantees. Its lane count is its pooled budget and the
cores held across all of its members can never exceed that budget. A group cannot
monopolize a machine by being split into more services, and an operator sizes a
tenant rather than each of its services separately.

\paragraph{Per-member ranking} Members are ranked for \emph{urgency} nonetheless
by their own active-thread count, not the group's. A member with no cores and a
packet waiting is considered congested and promptly served a core by the
allocator. Refused donations and fan-out fall back to ordinary park-and-wake,
which by definition cannot be donated a core.

\subsection{Allocator load}
\label{sec:design-cost}

Table~\ref{tab:cost} separates two things that are easy to mix up. One is what the
allocator spends \emph{looking} for work. The other is what a request actually
makes it \emph{do}.

\begin{table}[t]
  \caption{Allocator dataplane cost. $C$ managed cores, $A$ runtimes on the poll
    list, $K$ kthreads per runtime, $G$ groups, $H$ RPCs per request, $R$
    request rate.}
  \label{tab:cost}
  \small
  \begin{tabular*}{\columnwidth}{@{}@{\extracolsep{\fill}}>{\raggedright\arraybackslash}p{0.40\columnwidth}cc@{}}
    \toprule
     & Allocator-mediated & \sys \\
    \midrule
    \emph{polling} (fast pass, per iteration) & $\Theta(C)$
      & $\Theta(C)$ \\
    \emph{polling} (slow pass, per 10\,\us)  & $\Theta(A{\cdot}K)$
      & $\Theta(A{\cdot}K + G)$ \\
    \emph{work} (grants one request causes)  & $\Theta(H)$
      & $\Theta(1)$ \\
    \emph{work} (grants machine-wide)        & $\Theta(R{\cdot}H)$
      & $\Theta(R)$ \\
    \bottomrule
  \end{tabular*}
\end{table}

The two polling rows are almost unchanged. Reconciliation adds a constant per
core rather than a walk (\S\ref{sec:design-reconcile}), and the slow pass gains
only $\Theta(G)$ for the loopback sweep, dominated by the pre-existing
$\Theta(A{\cdot}K)$.

The two work rows are where the difference lives, and they are not measured by
the passes. They are what the passes \emph{initiate}. Placement by
observation parks the caller on both legs of every hop and needs an allocation
for each, $\Theta(H)$ per request, $\Theta(R{\cdot}H)$ machine-wide. \sys's
members hand the core over instead. A request over
\hotelres{}'s standard mix makes $\approx$5.6 RPCs and every leg
of each is a core transition, $\approx$9 of them between members which the group
takes care of. What remains for the allocator is the edge, and nothing that grows
with the depth of the call graph. \textbf{\sys's advantage is that the
allocator's load stops multiplying by hop count}, which is why the group's
latency is flat in call-graph depth and in tenant count where an
allocator-mediated one is not.

\section{Implementation}
\label{sec:impl}

\sys reuses the runtime and allocator of Caladan~\cite{fried2020caladan},
branched from its then-latest commit (\texttt{bdb4dde}). Caladan is
regularly maintained and is the scheduler of the newer
Junction~\cite{fried2024junction}; its allocator is the strongest published
instance of the placement-by-observation design of \S\ref{sec:bg-harvest}, and
its runtime already provides uthreads, kernel-bypass TCP/IP and the clean
preemption of a core that \S\ref{sec:design-revoke} rebuilds. Our changes span
the kernel module, allocator (the IOKernel) and runtime, adding about 10,500
lines and removing about 200; Table~\ref{tab:loc} in
Appendix~\ref{sec:app-loc} breaks that down. Two ioctls are the whole of the
kernel-module change. One is privileged, registering a group's threads, and the
other is the unprivileged \texttt{SWITCH\_TO} of \S\ref{sec:design-handoff}. The
IOKernel
gains a registration and region-allocation protocol on its control socket, a
reconciliation step in the per-core walk of every dataplane iteration, and
\S\ref{sec:design-policy}'s policy, which budgets a group as one entity while
ranking its members individually for urgency. In the runtime the changes
cluster at three points. The park path checks for a pending donation before it
parks, the send path decides whether a peer is reachable by loopback and arms a
donation, and \texttt{schedule()} handles first-hop placement and returns
lanes. Every path takes the fallback by ordinary control flow rather than
via an error handler. We apply a small set of fixes to stock Caladan before
measuring (Appendix~\ref{sec:app-baseline}); every Caladan number here includes
them.

\section{Evaluation}
\label{sec:eval}

\subsection{Methodology}
\label{sec:eval-method}

\paragraph{Testbed} One two-socket Intel Xeon Platinum 8360Y (Ice Lake, 36
cores per socket $\times$ 2 HT, 2.4\,GHz), 256\,GiB DRAM, Ubuntu 24.04 on Linux
6.8, Mellanox ConnectX-6 Dx. Runtimes, hugepages and
the NIC all sit on NUMA node 1. The IOKernel occupies that node's 72 hyperthreads,
keeps one physical core, and leaves 70 to allocate. Caladan runs in external
directpath with VFIO and EQ arming (MTU 9,000) using the default 25\,Gbps limit
for its bandwidth subcontroller; we disable the hyperthreading subcontroller,
which requires guaranteed physical cores and so admits only 35 services on our
35 cores, and whose symmetric guarantees cancel identically in IAS allocation
math anyway. We tune for low latency as recommended, disabling TurboBoost, CPU
idle states, frequency scaling and transparent
hugepages~\cite{leverich2014reconciling}. Ice Lake has no user interrupts, so
Caladan's UINTR path is off; it is Table~\ref{tab:wake}'s IPI+signal row, a
third of the tax, and does not touch the observation cost that grows with
tenant count (\S\ref{sec:eval-hop}).

\paragraph{Applications} We port DeathStarBench's
\hotelres~\cite{gan2019dsb} and Google's Online
Boutique~\cite{onlineboutique}, keeping each service graph, mix, paths, and
static data, and replacing gRPC/protobuf~\cite{grpc} with fixed-layout messages.
\hotelres{}'s memcached and MongoDB become an in-memory KV; Boutique has no
datastore. \hotelres is nine services and four levels deep, Boutique ten and
three. wrk2~\cite{wrk2} offers requests; locust~\cite{locust} offers
\emph{tasks}, two of which issue more than one HTTP request, so a load point of
30,000 is 30,000 RPS per tenant on \hotelres and 36,300 on Boutique. Rates we
report are in requests.

\paragraph{Best-effort antagonist} \texttt{x264\_be} runs 72 independent
single-stream x264 encoders~\cite{x264}, one per core, at BE priority.
Independence makes every reclaim a genuine preemption (a frame-threaded
encoder would yield at its sync points and collapse the tax into
Table~\ref{tab:wake}'s idle-core wake) and is the usual batch-transcode
configuration.

\paragraph{Deployment} A \emph{tenant} is one complete deployment, with its own
processes, address range and scheduling group; tenant count ranges over 2--10 (18--90
processes). Each tenant gets six lanes
(\S\ref{sec:design-invariant}). One load generator holds eight cores and
round-robins every tenant's frontend.

\paragraph{Arms and runs} Four arms. \texttt{Caladan} (with the
Appendix~\ref{sec:app-baseline} baseline fixes), \texttt{Caladan+tw} (idle
runtimes move from the poll list onto a timer wheel, the Junction variant),
\texttt{\sysname}, and \texttt{Linux} (batch task at \texttt{SCHED\_IDLE},
interior traffic over kernel loopback). In three cells (nine tenants at
30,000 RPS and ten at 25,000 and 30,000) the Linux generator can no longer
hold the arrival schedule and offers only 72--83\% of nominal; those three are
excluded from every range we quote for it. Every cell is three interleaved
repetitions of a 20\,s measurement and we plot the median of the three.
\sys's spread across repetitions is under 2\% of its median everywhere in the
grid; Caladan's is under 2\% from six tenants up and wider below five, where
its curve is steep enough that a run lands high or low on a slope rather than
on a level. We report \hotelres{}'s 10-RPC
\texttt{search} path (31\% of the mix) and Boutique's \browseprod (57\%, 12
RPCs), taking either per deployment and aggregating via a median.

\paragraph{Baseline provenance} The Caladan arms are not the 2020
artifact, but the maintained upstream tree at commit \texttt{bdb4dde}
(May 2026)~\cite{caladansrc}, which is the tree Junction~\cite{fried2024junction}
vendors as its scheduler and runtime~\cite{junctionsrc}. Junction makes
kernel bypass practical, with unmodified Linux binaries, a small host attack
surface, and thousands of instances per machine, but the allocator it makes
practical is the one measured here, and the two mechanisms it scales that
allocator with are in this build and in these arms
(\S\ref{sec:bg-remedies}). We do not run the services as uProcs in a single
instance. That is a different isolation boundary, and a different design point,
discussed in \S\ref{sec:related}.

\subsection{Latency}
\label{sec:eval-latency}

\begin{figure*}[t]
  \centering
  \includegraphics[width=\textwidth]{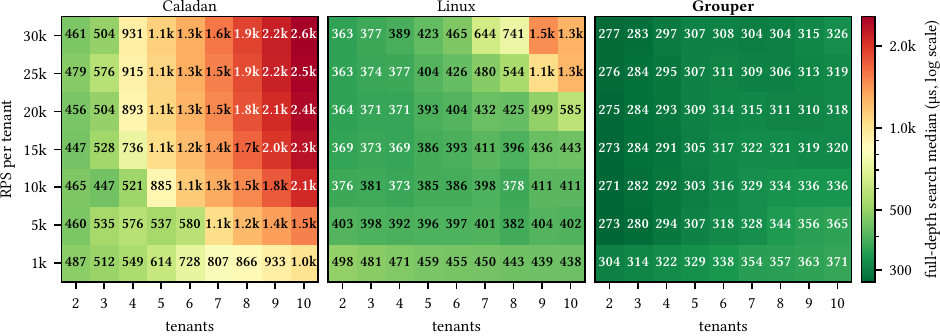}
  \caption{\textbf{Full-depth search median across the whole grid.} One panel per
    system; rows are load, columns are tenant count.}
  \label{fig:heat}
\end{figure*}

The grid crosses tenant count 2\dots\allowbreak10 with
1,000\dots\allowbreak30,000 requests per second per tenant, up to 300,000 in
total across 90 services on 72 cores against 72 encoders
(Figure~\ref{fig:heat}); Figure~\ref{fig:latency} took a cross-section at
20,000 RPS per tenant.

Caladan's search median rises from 456\,\us to 2,445\,\us over that line and its
99th percentile from 722\,\us to 3,539\,\us (5.4$\times$ and 4.9$\times$),
on a machine whose per-tenant load never varies. \sys goes from 275 to
318\,\us and from 407 to 891\,\us (1.2$\times$ and 2.2$\times$). The gap widens
monotonically in tenant count rather than remaining a constant factor, from
1.7$\times$ to 7.7$\times$ at the median and 1.8$\times$ to 4.0$\times$ at the
tail, which is the shape \S\ref{sec:design-cost} predicts. What a group removes
is $\Theta(R{\cdot}H)$ work at the allocator, and how much there is to remove
grows with the number of tenants asking.

\sys refuses 3.3--5.7\% of requests along this line and Caladan none; every
latency here is of a served request. Refusal is not what buys the gap. At ten
tenants no offered load in the grid puts Caladan where \sys is. Its median is
2,445\,\us at 20,000 RPS per tenant, 1,523\,\us at a quarter of that rate and
1,013\,\us at a twentieth, against \sys's 318\,\us at the full rate.
Appendix~\ref{sec:app-ablation} gives Caladan the same bound directly.

Linux is flat where Caladan is not, and its tail is why that is not enough.
With no central allocator to queue behind, its median rises only 364 to
585\,\us along the same line, beating Caladan at every tenant count, and
holds 363--1,057\,\us over the rest of the grid. Its 99th percentile does
not follow. It runs from 3,547 to 4,524\,\us, 3.3--11.0\,ms over the grid and
4.0--8.7$\times$ \sys's along this line,
because a latency-critical thread that wakes on a machine whose every core is
running the batch task waits out a scheduling slice it cannot
preempt~\cite{lozi2016wasted}. The slice is tunable and tuning it does not
close the gap. \texttt{base\_slice\_ns} defaults to 2.8\,ms here, a figure the
kernel scales with core count; sweeping it at two tenants takes the tail to a
floor of 1,958\,\us at 700\,\us and back up to 2,203 at 350, against
\sys's 407\,\us in the same cell. Its tail is
flat only because it is already high; \sys is the only arm that is low in
both.

Caladan+tw separates polling cost from allocation cost and confirms that the
latter is what matters. Unpolling idle runtimes is worth up to 43\% at the
lightest load in the grid, and more the more tenants there are to unpoll, but it
converges on stock as the machine fills. At ten tenants it is still within 13\%
of stock at the tail and 3.4$\times$ \sys. Making the scan cheaper delays the
cost by about an order of magnitude in offered rate; it does not remove it.

\subsection{Best-effort throughput}
\label{sec:eval-be}

\begin{figure}[t]
  \centering
  \includegraphics[width=\columnwidth]{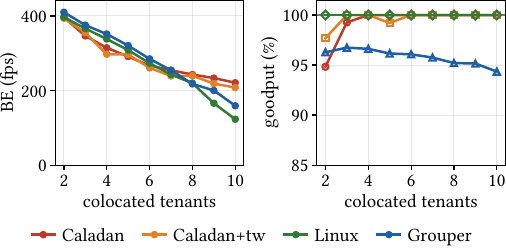}
  \caption{\textbf{What colocation costs}, vs.\ tenant count at 20,000 RPS per
    tenant. Best-effort throughput in frames/s on the left (idle baseline 523
    frames/s). Goodput as a percentage of offered load on the right, against the
    dashed offered line; the axis starts at 85\%.}
  \label{fig:be}
\end{figure}

\begin{figure}[t]
  \centering
  \includegraphics{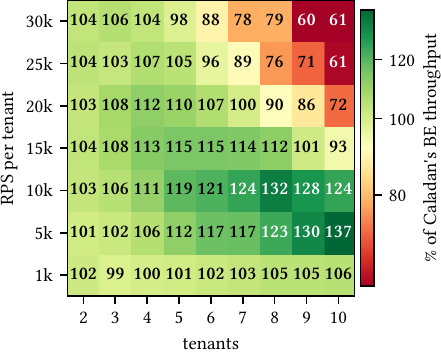}
  \caption{\textbf{Best-effort throughput across the whole grid.} \sys as a
    percentage of Caladan's, one cell per tenant count and load point.}
  \label{fig:beratio}
\end{figure}

Colocation is the reason these runtimes harvest cores at
all~\cite{lo2015heracles,iorgulescu2018perfiso}, so latency bought by starving
the batch task is not bought at all. Figure~\ref{fig:be} (left) takes the
20,000 RPS cross-section and Figure~\ref{fig:beratio} answers the question over
the whole grid. \sys leaves the encoders more throughput than Caladan
in
45 of 63 load points, by up to 37\%, within a tenth of it in 51, and less by at
most 40\% in the rest. Linux tracks Caladan along this line to seven tenants and
then falls away, leaving the encoders 123 frames/s at ten against Caladan's 221
and \sys's 159.

The mechanism is how often the batch task is disturbed. At ten tenants offering
20,000 RPS each, Caladan preempts the encoders 326{,}000 times a second
and \sys 29{,}400 (an eleventh as often). Cores are already busy, so the tax
is weakest here. At lower RPS more hops park and Caladan preempts considerably
more per request; \sys takes one core at the edge and carries it along the
call graph.

\subsection{Goodput}
\label{sec:eval-goodput}

Figure~\ref{fig:be} (right) shows \sys's achieved goodput at the 20,000 RPS
slice. At 1,000 requests per second per tenant \sys refuses 0.01\% and its
tail is already 1.5--2.3$\times$ better than Caladan's; at 5,000 it refuses at
most 0.09\% and is 1.6--3.0$\times$ better; at 10,000, under 1\%, it is
1.7--4.0$\times$ better. The advantage is there at essentially zero refusal, and
it grows with how much allocator work there is to remove, not with how much
traffic is turned away.

Linux refuses nothing and completes every request at all 63 load points, so its
line sits on the offered rate throughout, with Caladan's and Caladan+tw's
underneath it below four tenants. What that costs is the tail.
\sys's bound is on concurrency, not on rate. Goodput times mean residency is
what a group has in flight. At ten tenants (sixty lanes), \hotelres runs 2.2
and 20.7 requests in flight as the offered rate goes 1,000 and 10,000 per
tenant, refusing 0.0\% and 0.6\%; Online Boutique, whose requests are
2.12$\times$ as deep, runs 4.1 and 40.3 and refuses 0.0\% and 4.3\%. A deeper
request has a longer residency (324--400\,\us against 207--227), so Boutique
approaches the bound at a lower offered rate. Carrying one core along a path
bounds how many \emph{requests}
a group runs at once, never how many hops each makes
(Appendix~\ref{sec:app-ablation}).

\subsection{The cost of a hop}
\label{sec:eval-hop}

\begin{figure}[t]
  \centering
  \includegraphics[width=\columnwidth]{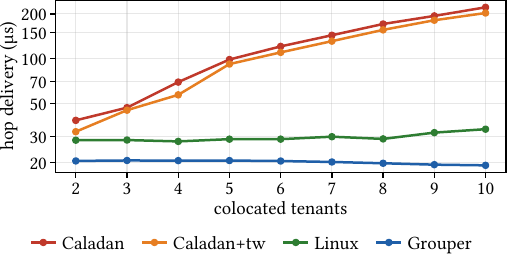}
  \caption{\textbf{Delivery time of one hop} vs.\ tenant count at 20,000 RPS
    per tenant. Median interior-hop latency, both legs, visit-weighted.}
  \label{fig:hopcost}
\end{figure}

\begin{table*}[t]
  \caption{Delivery time of one interior hop (\us) at equal total offered load,
    split across different numbers of tenants. Each row is one machine load;
    each half of it is the same requests per second arriving from more
    independent runtimes.}
  \label{tab:split}
  \begin{tabular*}{\textwidth}{@{}@{\extracolsep{\fill}}rlrrrlrrr@{}}
    \toprule
    total offered & split & Caladan & Linux & \sys & split & Caladan & Linux & \sys \\
    \midrule
     50,000 & 2$\times$25,000 &  38.5 & 28.1 & 20.6 & 10$\times$5,000  & 163.8 & 30.8 & 22.2 \\
     60,000 & 2$\times$30,000 &  38.2 & 27.9 & 20.6 & 6$\times$10,000  & 112.3 & 29.3 & 21.1 \\
     90,000 & 3$\times$30,000 &  46.5 & 28.2 & 20.6 & 9$\times$10,000  & 176.0 & 30.3 & 20.9 \\
    100,000 & 4$\times$25,000 &  74.2 & 27.8 & 20.6 & 10$\times$10,000 & 211.9 & 29.9 & 20.7 \\
    150,000 & 5$\times$30,000 & 101.2 & 29.2 & 20.5 & 10$\times$15,000 & 223.5 & 30.2 & 19.7 \\
    \bottomrule
  \end{tabular*}
\end{table*}

Figure~\ref{fig:hopcost} takes the same 20,000 RPS cross-section. Caladan's
interior hop costs 38.6\,\us at two tenants and 221.6 at ten (5.7$\times$), over
a sweep in which no tenant's rate, call graph, or per-hop work changes; \sys's
costs 20.6 and 19.2\,\us, and stays between 19.0 and 26.1 over the full grid.
Linux's hop is flat too, 28.4\,\us at two tenants and 33.6 at ten
(27.8--40.4 over the grid), which is the control that names the mechanism.
What stretches under Caladan is one allocator's reaction time, and a system
without one does not stretch.

What grows is the acquisition. Table~\ref{tab:split} holds total offered load
fixed and splits it across more tenants. Under Caladan a hop's delivery cost is
not a property of the tenant paying it but of how many independent runtimes are
asking at once, and a tenant that has changed nothing watches its own per-hop
cost quintuple because its neighbours exist. That coupling is what \sys
removes rather than mitigates.

\subsection{Doubling the hop count}
\label{sec:eval-depth}

\begin{figure}[t]
  \centering
  \includegraphics[width=\columnwidth]{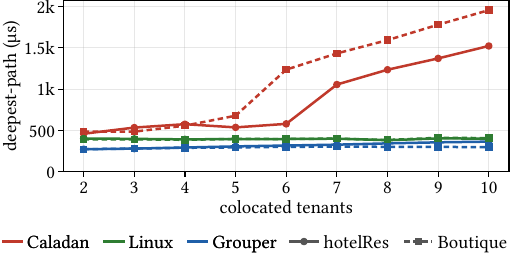}
  \caption{\textbf{Two applications, three systems} at 5,000 RPS per tenant.
    Deepest-path median for \hotelres (10 RPCs, solid) and Online
    Boutique (12, dashed). Load chosen so all arms complete $\geq$99.7\%.}
  \label{fig:depth}
\end{figure}

We ran the grid again against Online Boutique, at 2.12$\times$ the RPCs per
request, holding every other variable constant (Figure~\ref{fig:depth}). At ten
tenants an interior hop costs Caladan 163.8\,\us on \hotelres and 158.7 on
Online Boutique (the same price, set by how many runtimes are asking rather
than by which application asks), and a search request pays it nine times where a
browse request pays it eleven. The medians follow, 1,523 and 1,956\,\us.
\sys's hop stays roughly constant. Over the whole grid its median holds
271--371\,\us on the 10-RPC path and 273--337 on the 12-RPC one, with tails
of 370--1,019 and 375--1,318\,\us. Linux is flat too, its hop differing by
${\sim}$4\,\us between the two graphs and its median by none, 403 to 402\,\us
and 392 to 408. What separates those two flat systems is the tail. Linux sits at
3.5--4.0\,ms
for this cross-section against \sys's 375--1,318\,\us.

\subsection{The price of a reserved core}
\label{sec:eval-price}

\begin{figure*}[t]
  \centering
  \includegraphics[width=\textwidth]{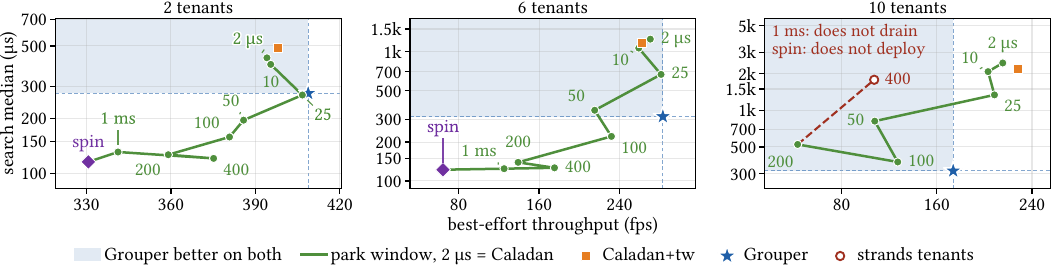}
  \caption{\textbf{What each remedy costs.} Search median vs.\ best-effort
    throughput at 20,000 RPS per tenant, one panel per tenant count, the park
    window swept and every setting labelled; shaded is where \sys wins on
    both axes. Each panel is fitted to its own data, so a position is
    comparable within a panel and not across them. At ten tenants 400\,\us
    strands 3 of 10 tenants.}
  \label{fig:cost}
\end{figure*}

\begin{figure}[t]
  \centering
  \includegraphics[width=\columnwidth]{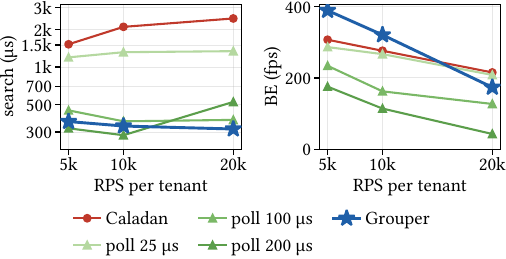}
  \caption{\textbf{The same comparison as offered load falls}, at ten tenants.
    Search median (left) and best-effort throughput (right) vs.\ RPS per
    tenant, on the same two axes as Figure~\ref{fig:cost}.}
  \label{fig:costload}
\end{figure}

\S\ref{sec:bg-remedies}'s remedies both reserve a core across the gap where the
request will need it again, charged to the batch task, by pinning a kthread per
service or widening the 2\,\us it spins before parking. We sweep that
whole window (Figure~\ref{fig:cost}). Widening it is free at first, and at 25\,\us
the search median falls 1.6--2.1$\times$ at every tenant count and the encoders
are no worse off up to eight tenants. But it is never enough, and eventually it
\textbf{fails}. At ten tenants 400\,\us strands 3 of 10 tenants, 1\,ms does not
drain, and spinning does not deploy. \sys holds 318\,\us while leaving
33\% of the batch machine, beating every setting from 50 to 200\,\us on both
axes; the three that leave the encoders more are 7.7$\times$, 6.6$\times$ and
4.2$\times$ slower, and at no tenant count is any setting both as fast as
\sys and as cheap. The tail is the one place a window competes, and at ten
tenants 100\,\us reaches 807\,\us against our 891, and pays a quarter of the
batch machine for it (24.3\% retained against 33\%) whether or not the load
that justified the reservation is still arriving. Spinning is the paper's
fastest arm (116\,\us, flat in
tenant count) and costs a core per service whether or not a request arrives, so
six tenants is a ceiling here with no load term in it. Nine microservices is
small, and \socialnet~\cite{gan2019dsb} would allow four tenants with its
datastores collapsed and two as it ships, so the remedy is least available
exactly where the tax it addresses is worst.

This cross-section is \sys's least favourable line, the only region of the
grid where its best-effort throughput falls below Caladan's
(Figure~\ref{fig:beratio}), and not where a peak-provisioned deployment spends
its time. Below it \sys recovers that axis
(Figure~\ref{fig:costload}): its best-effort deficit closes and then reverses
as load falls, while every park-window setting keeps the cost it paid at the
peak. At ten tenants it leaves the encoders 72\% of
Caladan's frames at 20,000 RPS, 124\% at 10,000 and 137\% at 5,000, holding the
median at 318, 336 and 365\,\us against Caladan's 2,445, 2,105 and 1,523. At
5,000 it leaves them more than any arm measured there, Caladan included, and at
none of the three loads is a window setting both faster and cheaper; Linux is
slower at all three. A reserved core's cost does not move with load, so
a bursty deployment pays the \emph{idle} figure (17\% of the batch machine at
two tenants, 42\% at six) almost all the time for a benefit only the peak
collects. \sys's cost is proportional to the work it does.

\section{Discussion and Limitations}
\label{sec:discussion}

\paragraph{The invariant is the system}
One core per request (\S\ref{sec:design-invariant}) rests on a synchronous
datapath and a concurrency bound. Appendix~\ref{sec:app-ablation} removes them
one at a time; neither carries the result. Keeping the handoff and sending
interior frames over the NIC
still closes 93\% of the median's distance from Caladan at ten tenants and 81\%
of the tail's; what the ring adds is that the frame and the core arrive
together, so what it closes grows with the machine, from 1.10$\times$ the tail
at two tenants to 1.56$\times$ at ten. Giving Caladan the bound instead does not
move it, and at \sys's own refusal rate it still costs 3.7--7.9$\times$ the
latency, because a bound limits how many requests pay the per-hop acquisition
and does not make a hop cheaper.

\paragraph{Fan-out} A group passes exactly one core along a path, yet
microservices fan out. \sys detects a branched request and donates to the
branch it sent for last; the rest fall back to park-and-wake, at a cost
Appendix~\ref{sec:app-fanout} measures. Production
analyses~\cite{luo2021alibaba} report that past a depth of two a tier holds a
single microservice with probability above 60\%, which is where single-core
handoff applies.

\paragraph{Asynchronous chains} Donation requires the sender to be done with
the core, which a blocking RPC is; an asynchronous send that leaves it runnable
falls back to park-and-wake (\S\ref{sec:design-group}). Previous studies find
that microservices typically issue nested RPCs and wait synchronously for the
results~\cite{pourhabibi2021cerebros}.

\paragraph{Sidecars} Service meshes~\cite{envoy,zhu2023meshinsight} put a
sidecar next to every service, which under Caladan doubles the boundary
crossings, and the parking tax, on every hop. Under \sys a sidecar is another
member and the handoff passes straight through it.

\paragraph{Portability} \sys changes the runtime scheduler, the kernel module
and the IOKernel without touching the programming model; applications are
unmodified apart from the optional hook that turns a flag into a
protocol-level refusal (\S\ref{sec:design-admission}). Built on Caladan, it
still requires applications to use that runtime, which
Junction~\cite{fried2024junction} does not, and a scheduling group would fit
there too.

\section{Related Work}
\label{sec:related}

\paragraph{Core-granting runtimes} \looseness=-1 Kernel-bypass
dataplanes~\cite{belay2014ix,prekas2017zygos,kaffes2019shinjuku,demoulin2021persephone,iyer2023concord},
userspace core
sizing~\cite{qin2018arachne} and the control-plane / data-plane
split~\cite{peter2014arrakis,marty2019snap} lead to
Shenango~\cite{ousterhout2019shenango} and Caladan~\cite{fried2020caladan} and
their privileged microsecond-granularity allocator, which
Junction~\cite{fried2024junction} carries forward and scales to thousands of
instances. All schedule \emph{one application's} requests onto \emph{that
application's} cores, and all require the allocator to first \emph{discover}
that a core should move. \sys keeps that machinery and removes it only from
the per-hop critical path, within a trusted set of runtimes.

\paragraph{Making core movement faster} \looseness=-1 A parallel line makes the transfer
itself cheap (Vessel~\cite{lin2024vessel} with MPK domains and user interrupts,
HyperFlux~\cite{yan2026hyperflux} between lightweight VMs,
Junction~\cite{fried2024junction} with uProcs in one address space). Running a
tenant's services as uProcs removes the cost measured here outright, by making
the tenant a single scheduling unit with no boundary left to cross; a group
makes the tenant a single allocation unit too, without merging the address
spaces. Nu~\cite{ruan2023nu} moves the \emph{work} instead, and Service
Weaver~\cite{ghemawat2023weaver} erases the boundary altogether. Each pays with
the process boundary, a domain limit, or hardware such as
UINTR~\cite{intelsdm,stojkovic2025hardharvest} commodity servers lack, making
reallocation fast where a group makes the decision free. Collapsing the
boundary is available to any deployment willing to give up the decomposition; a
group is what is available when it is not.

\paragraph{Cross-domain control transfer} Directed handoff has a long lineage.
LRPC~\cite{bershad1990lrpc}, URPC~\cite{bershad1991urpc},
L4~\cite{liedtke1993ipc}, Mach~\cite{black1990mach} and seL4's scheduling
contexts~\cite{lyons2018scheduling} all lend a thread, or the right to run,
across a domain boundary~\cite{mi2019skybridge,du2019xpc}, but always inside a
kernel that was itself the sole scheduler. Here a preemptive userspace
allocator owns the cores, so keeping its accounting exact
(\S\ref{sec:design-reconcile}) and revoking a core whose occupant it does not
know (\S\ref{sec:design-revoke}) is a different problem. It is not a futex
swap~\cite{umcg}, which moves no core, nor a
ghOSt~\cite{humphries2021ghost} or Syrup~\cite{kaffes2021syrup} policy, which
cannot beat a round trip to the agent.

\paragraph{Microservice fast paths} Nightcore~\cite{jia2021nightcore},
SPRIGHT~\cite{qi2022spright}, SAND~\cite{akkus2018sand} and service
meshes~\cite{envoy} accelerate the \emph{messages} between colocated
services~\cite{li2019socksdirect}, and \uT~\cite{sriraman2018utune} shows how
much of an OLDI tail per-hop threading sets. All are complementary, and we build
one ourselves (\S\ref{sec:design-loopback}), but they move the bytes while the
receiver stays parked. Delivering the message \emph{with} the core is what
removes the wakeup.

\paragraph{Coscheduling and overload control}
\looseness=-1
Coscheduling~\cite{ousterhout1982coscheduling,arpacidusseau2001implicit,feitelson1992gang}
runs
communicating processes \emph{simultaneously}; a chain is the opposite, and
co-residency wastes cores in proportion to its length
(\S\ref{sec:eval-price}). Callisto~\cite{harris2014callisto,tucker1989process} and
Akaros~\cite{rhoden2011akaros} make granted cores
first-class~\cite{colmenares2013tessellation}; a group stretches
that across process boundaries. Breakwater~\cite{cho2020breakwater} and related
schemes~\cite{zhou2018wechat,welsh2001seda,cho2023protego} regulate overload
with RPC-granularity credits
(\S\ref{sec:design-admission}).

\section{Conclusion}
\label{sec:concl}

Kernel-bypass runtimes made microsecond-scale colocation practical by harvesting
cores aggressively, and that made wakeups expensive. Microservice chains pay
that tax on every hop in both directions, and on a shared host it compounds,
since every colocated tenant's hops arrive at the same central allocator.

The information needed to move a core is in the application's dataflow.
A scheduling group lets mutually trusting services act on it, handing cores
through an unprivileged kernel fast path while the allocator keeps control
through reconciliation, core-addressed revocation and a pooled budget.
Allocator load falls from $\Theta(R{\cdot}H)$ to $\Theta(R)$, median nearly
flat and tail under a millisecond as tenants and load scale, each service
still in its own address space, failure domain and release cycle.

\bibliographystyle{ACM-Reference-Format}
\bibliography{refs}

\appendix

\section{Implementation size}
\label{sec:app-loc}

Table~\ref{tab:loc} counts lines added and lines removed against the Caladan branch
point from which \sys was developed (commit \texttt{bdb4dde}).

\begin{table}[ht]
  \caption{Implementation size (lines added / removed against the branch
    point).}
  \label{tab:loc}
  \small
  \begin{tabular*}{\columnwidth}{@{}@{\extracolsep{\fill}}>{\raggedright\arraybackslash}p{0.60\columnwidth}rr@{}}
    \toprule
    component & added & removed \\
    \midrule
    \texttt{ksched} kernel module & 391 & 1 \\
    IOKernel (control, \texttt{sched.c}, \texttt{ias.c},
      directpath, stats)          & 4,181 & 83 \\
    runtime scheduler, kthreads, net stack & 3,609 & 111 \\
    runtime loopback datapath (\texttt{sched\_group\_lb.c}) & 975 & 0 \\
    shared headers (\texttt{inc/}, \texttt{base/}) & 1,403 & 8 \\
    \midrule
    \textbf{total} & \textbf{10,559} & \textbf{203} \\
    \bottomrule
  \end{tabular*}
\end{table}

\section{Fixes to the baseline}
\label{sec:app-baseline}

Stock Caladan was taken at its latest commit at the time (\texttt{bdb4dde}).
Several defects were fixed prior to measuring, and every Caladan number in this
paper includes them. The primary fix was in the IOKernel's idle fast-wake path,
which checks for arriving packets between periodic allocation passes. Stock
Caladan evaluated arrivals by calculating descriptor age against a device
clock refreshed only in the slow pass; newly arrived packets registered as zero
delay and missed immediate wakeups. Testing the descriptor's completion parity
bit directly detects work without clock overhead, restoring prompt wakeups.
The remaining fixes address concurrency races and buffer exhaustion under heavy
multi-tenant load. Publication races during TCP connection setup caused
spurious timeouts, stalled receive pollers and buffer pool deadlocks from
deferred transmit completions and pinned receive strides. With these fixes,
stock Caladan runs stably, without artificial stalls or drops.

\section{Ablations}
\label{sec:app-ablation}

The invariant of \S\ref{sec:design-invariant} rests on two mechanisms. Removing
either leaves most of \sys's advantage intact. The bound's size is a third
question of how many lanes and what that spends, not a third mechanism.
All three experiments keep \S\ref{sec:eval-method}'s configuration otherwise, at
three interleaved repetitions.

\subsection{The shared-memory datapath}
\label{sec:app-abl-datapath}

\begin{figure}[t]
  \centering
  \includegraphics[width=\columnwidth]{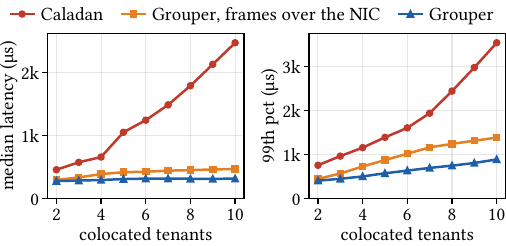}
  \caption{\textbf{What the ring is worth.} Full-depth \texttt{search} median
    (left) and 99th percentile (right) at 20,000 RPS per tenant.
    \texttt{groupwire} is \sys with interior frames sent over the NIC
    instead of through the peer's ring; nothing else about that group
    changes.}
  \label{fig:datapath}
\end{figure}

\texttt{groupwire} keeps the scheduling group entire (the handoff, the lanes,
the pinned receive queues, the pooled budget) and sends interior frames over
the NIC, naming the destination pad in the IP header so that a frame still
lands on the kthread its sender donated to.

Most of \sys remains without the ring (Figure~\ref{fig:datapath}). At ten
tenants the handoff alone closes
93\% of the median's distance from Caladan and 81\% of the tail's. What it does
not close grows with the machine. The wire arm's tail is 1.10$\times$ the
ring's at two tenants and 1.56$\times$ at ten, because the frame and the core
are then dispatched separately and \emph{drift}. The request waits for whichever
of the two is later. Moving the bytes is worth about a microsecond; moving them \emph{with}
the core is worth the rest.

\subsection{The concurrency bound}
\label{sec:app-abl-admission}

\begin{figure}[t]
  \centering
  \includegraphics[width=\columnwidth]{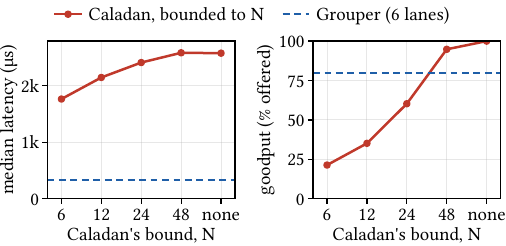}
  \caption{\textbf{No bound places Caladan where \sys is.} Stock Caladan
    bounded to $N$ requests in flight per deployment, at ten tenants and
    30,000 RPS each; $N{=}6$ is \sys's own lane count and \texttt{none} is
    unbounded Caladan. At no $N$ is Caladan below \sys's level on the left
    \emph{and} above it on the right.}
  \label{fig:admission}
\end{figure}

\sys runs one request per lane and refuses the rest at the edge, so a
question \S\ref{sec:eval} leaves open is where Caladan then lands when it is bounded
the same way. \texttt{cap}$N$ is stock Caladan bounded to $N$ requests in flight
per deployment, refused at the frontend with the same reply a lane denial
produces.

No bound gets Caladan there (Figure~\ref{fig:admission}). At six permits
against six lanes (ten tenants,
30,000 RPS), the two run the same concurrency, 5.8 requests in flight against
5.5, and Caladan is 5.4$\times$ slower at the median while refusing 78.6\%
against 20.0\%; at \sys's own refusal rate it costs 3.7--7.9$\times$ the
latency across the grid. Little's law says why, from residency alone.
An interior hop costs \texttt{cap6} 183.9\,\us against 19.1, which over
${\approx}$5.6 RPCs is essentially the whole of Caladan's 912\,\us residency
against 230. The same six permits carry four times the traffic because what
occupies them is 9.6$\times$ cheaper. Three of that arm's effects favour
Caladan. A bounded deployment loads the machine less, so its own hop gets
13--29\% cheaper; a refusal is issued when the system is momentarily full, so a
tight cap serves requests drawn from calmer instants; and \texttt{cap6} leaves
the encoders more frames than \sys while delivering a fifth of the
traffic.

\subsection{The lane count}
\label{sec:app-abl-lanes}

\begin{figure}[t]
  \centering
  \includegraphics[width=\columnwidth]{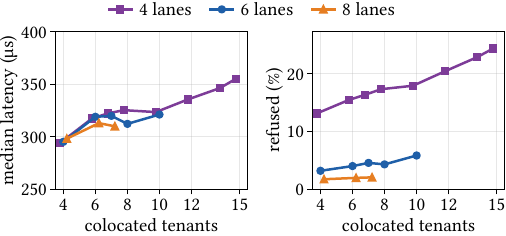}
  \caption{\textbf{The hop does not shift with the bound.} Full-depth search
    median (left) and fraction refused (right) at 20,000 RPS per tenant. A
    line that continues further right packs more tenants. Six is
    \S\ref{sec:eval-method}'s setting.}
  \label{fig:lanes}
\end{figure}

A request holds one lane for its residency, and there are as many lanes as a
member has kthreads, so the lane count is the group's concurrency bound and
its peak core claim together (\S\ref{sec:design-invariant}).
\S\ref{sec:eval} measures one setting of that knob (six) and
Appendix~\ref{sec:app-abl-admission} asks whether the bound is the result.
This asks what that six is (Figure~\ref{fig:lanes}).

\sys only, at \S\ref{sec:eval}'s 20,000 RPS cross-section, where six lanes
refuse 3.3--5.7\%. A cell that would book more cores at peak occupancy than the
machine has, after the spinning load generator, is not run. The IOKernel leaves
70 to allocate, eight of them guaranteed to the generator, 62 left for LC, so
four lanes reach 15 tenants, six reach 10 and eight reach 7.

The hop itself does not move (Figure~\ref{fig:lanes}, left). At 7 tenants
the full-depth search median is 322, 320 and 310\,\us at four,
six and eight lanes. What moves is the bound (right). Four lanes refuse
16.3\%, six 4.6\%, eight 2.1\%. The same four lanes at 15
tenants (a machine six cannot occupy) refuse 24.3\%
and serve 227{,}000 requests a second against eight lanes at 7 tenants
serving 137{,}000 and refusing 2.1\%. Fewer lanes pack more tenants and
refuse more of each; more lanes do the reverse. Six is a point on that tradeoff.

\section{Fan-out}
\label{sec:app-fanout}

\begin{figure}[t]
  \centering
  \includegraphics[width=\columnwidth]{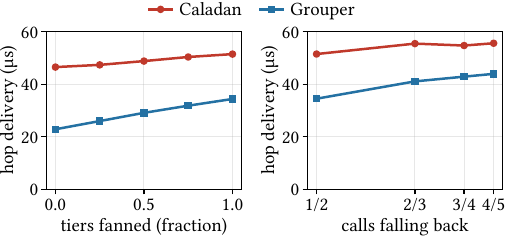}
  \caption{\textbf{Fan-out costs \sys its donations, not its advantage.}
    Delivery time of one interior hop. How many of the four tiers fan on the
    left, at degree two. How wide every tier fans on the right, as the fraction of a
    tier's calls that fall back onto park-and-wake.}
  \label{fig:fanout}
\end{figure}

A tier of degree $R$ spends $R$ donations when it calls serially and one when
it fans, so $R-1$ of $R$ fall back to ordinary park-and-wake. Figure~\ref{fig:fanout} sweeps both
axes on a synthetic graph of four tiers over nine services at a fixed depth of
five, varying how many tiers fan, at degree two, and how wide every tier fans. A hop
costs 22.8\,\us with nothing fanning, 34.4\,\us with all four tiers fanned, and
43.9\,\us at degree five, where four calls in five fall back to ordinary park-and-wake, against Caladan's
46.5--55.6\,\us, unmoved because it pays an allocator wake either way. The cost
rises and then saturates without crossing. The chain call and every reply leg are
still handed over directly, whatever the width. Detection is exact, firing 1.02
times per request per fanned tier and 0.04 times on the sequential graph, and
the fallback costs the batch task nothing extra. Best-effort throughput stays within
2\% of Caladan's at every point on both axes.

The settings match \S\ref{sec:eval-method}'s but for three. They use two tenants at 2,000
RPS each over twelve lanes a service, rather than ten at 20,000 over six. A
fanned request holds its lane until its widest branch returns, so at
\S\ref{sec:eval}'s cross-section the refusal rate would rise with the fanned
fraction and the $x$ axis would be admission control rather than fan-out; here
the group refuses under 0.04\% of requests at each point.

\end{document}